\RequirePackage{fix-cm}
\documentclass[smallextended]{svjour3}       
\smartqed  
\usepackage{url,graphicx,graphics,amssymb,amsmath,relsize,supertabular}
\usepackage[pdfstartview={FitH},pdffitwindow=true,colorlinks=true,citecolor=blue,linkcolor=blue]{hyperref}%
\usepackage{array,longtable}
\usepackage{braket}
\usepackage{mathptmx}      
\usepackage{latexsym}
\begin{document}

\title{Quantum Belief Propagation Algorithm versus Suzuki-Trotter approach in the one-dimensional Heisenberg chains
}


\author{Farzad Ghafari Jouneghani         \and
        Mohammad Babazadeh           \and
        Davoud Salami\and
        Hossein Movla 
}


\institute{Farzad Ghafari Jouneghani, Mohammad Babazadeh, and Davoud Salami \at
              Faculty of Physics, University of Tabriz, Tabriz, Iran \\
              Tel.: +989398990564\\
              \email{f.ghafari89@ms.tabrizu.ac.ir}           
\and
          Hossein Movla, \at
              Azar Aytash Co., Technology Incubator, University of Tabriz, Tabriz, Iran
}

\date{Received: date / Accepted: date}

\maketitle

\begin{abstract}
Quantum systems are the future candidates for computers and information processing devices. Information about quantum states and processes may be incomplete and scattered in these systems. We use a quantum version of Belief Propagation(BP) Algorithm to integrate the distributed information. In this algorithm the distributed information, which is in the form of density matrix, can be approximated to local structures. The validity of this algorithm is measured in comparison with Suzuki-Trotter(ST) method, using simulated information. ST in 3-body Heisenberg example gives a more accurate answer, however Quantum Belief Propagation (QBP) runs faster based on complexity. In order to develop it in the future, we should be looking for ways to increase the accuracy of QBP .
\end{abstract}

\section{Introduction}
\label{intro}
In 1988, Pearl designed BP algorithm to solve marginalization and other inference problems. BP algorithm is an approximate message passing algorithm which uses a graph illustration called graphical models to solve problems \cite{Pearl1}. On the other hand, in 1963, Gallagher had designed a decoding algorithm similar to BP algorithm for decoding his known and valuable code, Low Density Parity Check \cite{Gallager2}. One of the important points in functionality and generality of BP algorithm is that it appears in various fields of sciences concerned with mathematics. Theory of codes \cite{Gallager2,McEliece3}, physics \cite{MacKay4,Yedidia5}, statistics, and artificial intelligence \cite{Pearl1,Kindermann6} can be noted in this regard. Unfortunately, this diversity has caused the algorithm to lack a standard fully unified notation. So many generalizations, from basic messenger algorithms to very complex approaches, have been done on BP so far.
BP can be used in graphical models such as Markov Random Field \cite{Kindermann6}, Bayesian Networks \cite{Ben7}, and Factor Graph \cite{Kschischang8} with an almost similar notation. Recently, significant progresses have been done in the field of practical application of BP algorithm for loopy graphs. One of the other areas in which this algorithm illustrate a good performance is satisfiability problems. It is known as Survey Propagation which can solve very difficult satisfiability problems \cite{Braunstein9}. In this paper an attempt is made to review the basic principles of the BP algorithm. These principles can help solve problems based on the transmission of messages.
Generalizing the graphical models and related algorithms to quantum mechanics has recently attracted researchers’ attention. The first and most fundamental step in this regard was taken by Leifer et al. in 2008 \cite{Matt10}. Later on, its application in quantum statistical mechanics \cite {Poulin11,Hasting12,Bilgin13} and quantum codes \cite{Pawel14} were investigated and the acceptable results were obtained. One of the main reasons for this generalization is nature of quantum mechanics as a probabilistic theory with an origin exactly the same as graphical models. Despite these basic similarities, differences between quantum mechanics and classical probability theory make this generalization complicated. Therefore, establishing a unique and worthy theoretical framework, in which we can put quantum mechanics and graphical models, can be very challenging.
The theoretical foundation of classical and QBP algorithm is developed in Sec. 1 and 2 and then ST approximation was discussed in Sec. 3. At the end, in Sec. 4 the performance of these approximation methods was compared in an example of statistical physics. First of all, we were trying to have an overview of previous researches. After that, we hope to explore the strengths and weaknesses of each of the proposed methods by solving simple problems.
\\
\section{Marginal Probability}
\label{sec:1}
Calculation the marginal probabilities is one of the most challenging problems in probability theory \cite{Trumpler15}. We want to calculate this probability such that a random variable in a probability distribution function selects one of the possible random states. This probability is called marginal probability of the relevant variable. Mathematically, the marginal probability is defined by summation over all possible states of all variables except the target variable. For example, if we have the total probability distribution $p(x_1,x_2,...,x_N)$ and we want to calculate the marginal probability for nth random variable, we have:\\
\begin{equation}
p(x_N)=\sum_{x_1}\sum_{x_2}...\sum_{x_{N-1}}p(x_1,x_2,...,x_N)
\end{equation}
Marginal calculation for large distributions directly and by using Formula 1 is very complex and virtually impossible. What will be introduced later as belief is approximate marginal calculation.\\
\section{Belief Propagation Algorithm (BP) \cite{Jonathan16} }
\label{sec:2}
Arbitrary probability distribution $p ({x})$ is used in order to introduce the BP algorithm. Main focus is on $x_{i}$ random variables which are entered in the total probability distribution $p(\{x\})$ as the following (taking into account the distribution like this does not detract from the totality of the problem).\\
\begin{equation}
p(\{x\})=\frac{1}{z}\prod_{ij}\Psi_{ij}(x_i,y_j)\prod_{i}\Phi_{i}(x_i)
\end{equation}\\
$\Phi_{i}(x_i)$ and $\Psi_{ij}(x_i,y_j)$ factor the total probability function and $Z$ is a normalization constant.\\
Variables such as $m_{ij}(x_j)$ are introduced in BP algorithm. These new variables can be introduced implicitly as a "message" from random variable $i$ to random variable $j$. This message indicates that variable $j$ should be placed in which state (see Fig. 1), where $m_{ij}(x_j)$ will be a vector with the same dimensions as $x_{j}$\'s dimensions (states that can be taken by $x_{j}$ ). If $ith$ variable thinks that $jth$ variable is in its $kth$ state with $p$ probability, then $k$th components of the vector $m_{ij}(x_j)$ is proportional to the $p$ probability.\\
In BP algorithm, belief in $i$ th variable is proportional to the product of the local function and all incoming messages to $i$:
\begin{equation}
b_{i}(x_{i})=Z \Phi_i(x_i) \prod_{j\in N(i)}m_{ji}(x_i)
\end{equation}
Where $Z$ is a normalization constant (the sum of Beliefs must be equal to 1) and $N(i)$ represents the neighbors of $i$. Messages are updated in a self-compatible manner by the following Eq.:
\begin{equation}
m_{ij}(x_j)=\sum_{x_i} \Phi_i(x_i)\Psi_{ij}(x_i,x_j) \prod_{k\in N(i)/j}m_{ki}(x_i)
\end{equation}
$N(i)/j$ is the set of all neighbors of $i$ except $j$. Marginal probability of two points $i$, $j$ is introduced where $i$ and $j$ are neighboring points, in which the marginal probability is given by marginalization of total probability distribution function on all points except $i$ and $j$.
\begin{figure}
\includegraphics[width=0.6\textwidth]{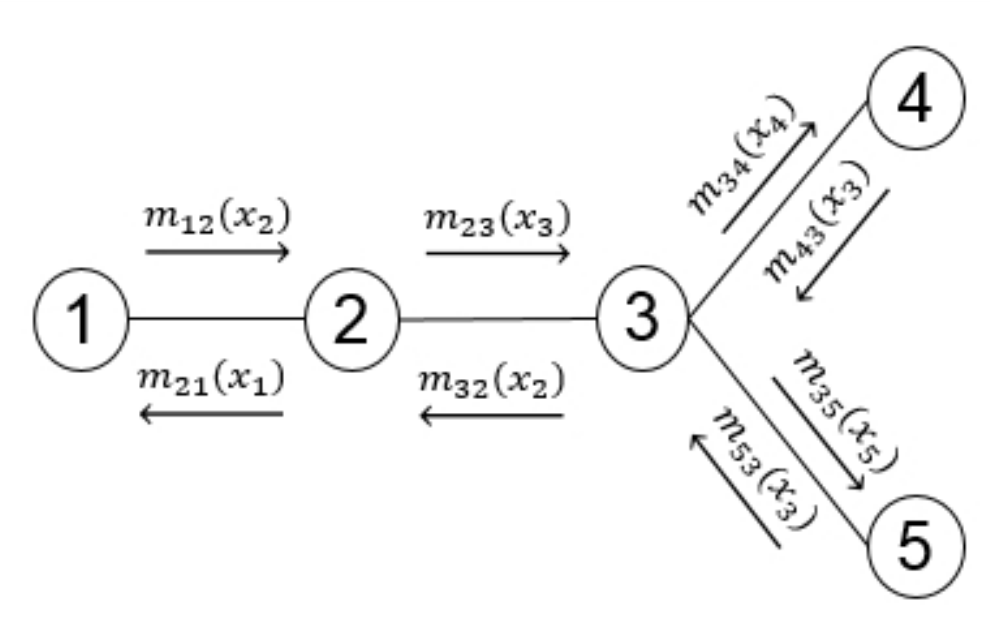}
\caption{Schematic illustration of exchanged messages between nodes of graph}
\label{fig:1}       
\end{figure}
\begin{equation}
p_{ij}(x_i,x_j)=\sum_{\{x\}\/x_{i},x_{j}} p(\{x\})
\end{equation}\\
BP for the two-point Belief $b_{ij}(x_i,x_j)$ is written similarly as Eq.$(3)$ for single-point Belief as follows:\\
\begin{equation}
b_{ij}(x_i,x_j)= K \Psi_{ij}(x_i,x_j)\phi_i(x_i)\phi_j(x_j)\prod_{k\in N(i)/j}m_{ki}(x_j)\prod_{l\in N(j)/i}m_{lj}(x_j)
\end{equation}
Where $K\in N(i)/j$ means all neighboring points, except point $j$.
\section{Quantum Belief Propagation}
The trend that we are following in quantum part \cite{Matt10,Tanaka17} is quite similar to the classical trend but there is a difference between them. Probability distribution function and marginal probabilities in the classical approach are respectively replaced by Density matrix and reduced matrices in QBP.Therefore, by quantum belief we mean approximate calculation of the reduced matrices. What we are dealing with here and  specifically intend to apply it,  is density matrix.
\begin{equation}
\rho(v)=\frac{\exp((-H(V))}{tr\exp((-H(V))}
\end{equation}
\begin{equation}
H(V)= \sum_{\sigma\in V}H(\sigma)
\end{equation}
\begin{equation}
\bra{\overrightarrow{\vec{x}}}H(\sigma)\ket{\overrightarrow{\vec{y}}}=\bra{\overrightarrow{\vec{x_\sigma}}}E(\sigma)\ket{\overrightarrow{\vec{y_\sigma}}}\prod_{i\in V\setminus\sigma}\delta_{x_i,y_i}
\end{equation}
In Eq.(8), V is the set of vertices and Hamiltonian is expressed in terms of smaller terms. For example, in Heisenberg model, that we will review it later, the total Hamiltonian is expressed in terms of binary interactive statements that will result in $\mid \sigma \mid=2$. Therefore, the reduced density matrix for a set of points or single points can be defined as follows:
\begin{equation}
\rho(\sigma)=tr_{\backslash i}\rho(v)
\end{equation}
\begin{equation}
\rho(i)=tr_{\backslash\sigma}\rho(v)
\end{equation}
By applying normalization and marginalization conditions $tr(\rho(i))=1$, $tr(\rho(\sigma))=1$, and $tr_{\backslash i}\rho(\sigma)=\rho(i)$ exactly like in classic, the following Eq.s are obtained for beliefs ($Q$s are represented for Beliefs)for single and paired points, and the messages updating rules (given $\mid \sigma \mid=2$ , that doesn’t detracts the totality).
\begin{equation}
Q_{i}\cong\frac{1}{Z_i}\exp\lgroup\sum_{k\in N_i}m_{k\rightarrow i}\rgroup
\end{equation}
\begin{equation}
Q_{ij}\cong\frac{1}{Z_{ij}}\exp\lgroup -E_{ij}+\sum_{k\in N_{i}\backslash j}m_{k\rightarrow i}\otimes I+\sum_{l\in N_{j}\backslash i}I\otimes m_{l\rightarrow j}\rgroup
\end{equation}
\begin{align}
&m_{j\rightarrow i}=-\sum_{k\in N_{i}\backslash j}m_{k\rightarrow i} +  \log \lgroup\frac{Z_i}{Z_{ij}}tr_{\backslash i}\lgroup \exp \lgroup -E_{ij}+ \sum_{k\in N_{i}\backslash j}m_{k\rightarrow i}\otimes I+\nonumber\\
& \sum_{l\in N_{j}\backslash i}I\otimes m_{l\rightarrow i}\rgroup\rgroup\rgroup
\end{align}
Z’s are normalization constants and I is the unit matrix.\\
\section{Suzuki-Trotter semi-classical approximation:}
A semi-classical approximation which is known as ST \cite{Suzuki18} is introduced in this section. To better understand this method, we use a convenient Hamiltonian in which the internal statements cannot commute with each others.\\
For example, if Hamiltonian, which is $H=h_{12}+h_{23}$ provided that  $[h_{12},h_{23}]\neq0$ describeS a 3-spin system, we first convert the problem into classical mode using Suzuki –Trotter formula \cite{Suzuki18} to calculate the density matrix \cite{Pathria19}, then it will be solved using classical algorithm.
\begin{align}
&\exp(-h_{12}-h_{23})=\lim_{n \to +\infty} \lgroup\exp\lgroup\frac{-1}{n}h_{12}\rgroup\times
\exp\lgroup\frac{-1}{n}h_{23}\rgroup\rgroup^n\nonumber\\&    \text{(\emph{Suzuki--Trotter formula})}
\label{Eq.15}
\end{align}
\begin{equation}
\rho=\frac{\exp(-\beta H)}{tr (\exp (-\beta H))}
\end{equation}
To solve this problem, we first need to convert ρ into a classical distribution and introduce the classical variables and then obtain ρ matrix elements using classical algorithm. Therefore, one more time we repeat ST calculations.
\begin{align}
&\rho=\exp(-\beta h_{12}-\beta(h_{23})=\lim_{n \to +\infty}\sum_{\{\vec{c_1},\vec{c_2},...,\vec{c_{n-1}}\}}
\nonumber\\
&\ket{\vec{a}}\bra{\vec{a}}
\exp \lgroup -\frac{\beta}{n}h_{12}
\rgroup \exp \lgroup -\frac{\beta}{n}h_{23}\rgroup\nonumber\\
&\ket{\vec{c_1}}\times\bra{\vec{c_1}}
\exp \lgroup -\frac{\beta}{n}h_{12}\rgroup\exp \lgroup -\frac{\beta}{n}h_{23}\rgroup\nonumber\\
&\ket{\vec{c_2}}\times\bra{\vec{c_2}}\exp \lgroup -\frac{\beta}{n}h_{12}\rgroup\exp \lgroup -\frac{\beta}{n}h_{23}\rgroup\nonumber\\
&\ket{\vec{c_3}}\times
\ldots\times\bra{\vec{c_n-1}}\exp \lgroup -\frac{\beta}{n}h_{12}\rgroup\exp \lgroup -\frac{\beta}{n}h_{23}\rgroup\nonumber\\
&\Braket{\vec{b}|\vec{b}}=
\lim_{n\to+\infty}\sum_{\{\vec{c_1},\vec{c_2},...,\vec{c_{n-1}}\}}
\ket{\vec{a}}
W(\vec{a};\vec{c_1})\nonumber\\
&W(\vec{c_1};\vec{c_2})W(\vec{c_2};\vec{c_3})\times\ldots\times W(\vec{c_{n-1}};\vec{b})\bra{\vec{b}}
\end{align}
 The above Eq. can be rewritten as follows:
\begin{subequations}
\begin{equation}
\rho=\frac{\exp(-h_{12}-h_{23})}{tr\lgroup\exp(-h_{12}-h_{23})\rgroup}
\end{equation}
\begin{equation}
=\sum_{\vec{a}}\sum_{\vec{b}}\ket{\vec{a}}\lgroup\lim_{n \to +\infty}\sum_{\{\vec{c_1},\vec{c_2},...,\vec{c_{n-1}}\}}P(\vec{a},\vec{c_1},\vec{c_2},...,\vec{c_{n-1}},\vec{b})\rgroup\bra{\vec{b}}
\end{equation}
\end{subequations}
where
\begin{align}
&P(\vec{a},\vec{c_1},\vec{c_2},...,\vec{c_{n-1}},\vec{b})=\nonumber\\
&\frac{W(\vec{a};\vec{c_1})W(\vec{c_1};\vec{c_2})\times\ldots\times W(\vec{c_{n-2}};\vec{c_{n-1}})W(\vec{c_{n-1}};\vec{b})}{\sum_{\vec{a},\vec{c_1},\vec{c_2},...,\vec{c_{n-1}},\vec{b}}W(\vec{a};\vec{c_1})W(\vec{c_1};\vec{c_2})\times\ldots\times W(\vec{c_{n-2}};\vec{c_{n-1}})W(\vec{c_{n-1}};\vec{b})}
\end{align}
To calculate the matrix $\rho$, we reached $P$ calculation which is a quite classical distribution and a function of the variables $a,c_1,c_2,...,c_{n-1},b$. Actually $P$ is factored in terms of  $W$ functions. Now we have all the necessary tools for solving a classical BP problem Calculation of element $(a, b)$ in density matrix is equivalent to calculation of marginal probability $P(a, b)$. Therefore, we must apply the BP to calculate the messages. If the normalization constant is shown by $Z$, then message calculation would be as follows:
\begin{subequations}
\begin{align}
&P(\vec{a},\vec{c_1},\vec{c_2},...,\vec{c_{n-1}},\vec{b})=\frac{1}{Z}W(\vec{a};\vec{c_1})W(\vec{c_1};\vec{c_2})\times\ldots\times\nonumber\\
&W(\vec{c_{n-2}};\vec{c_{n-1}})W(\vec{c_{n-1}};\vec{b})
\end{align}
\begin{align}
&P(\vec{a},\vec{c_1},\vec{c_2},...,\vec{c_{n-1}},\vec{b})=\frac{1}{Z}W(\vec{a};\vec{c_1})W(\vec{c_1};\vec{c_2})\times\ldots\times\nonumber\\
&\underbrace{ \sum_{\vec{c_{n-1}}}W(\vec{c_{n-2}};\vec{c_{n-1}})W(\vec{c_{n-1}};\vec{b})}_{m_{n-1 \rightarrow n-2}(\vec{c_{n-2}})}
\end{align}
\begin{align}
&P(\vec{a},\vec{c_1},\vec{c_2},...,\vec{c_{n-1}},\vec{b})=\frac{1}{Z}W(\vec{a};\vec{c_1})W(\vec{c_1};\vec{c_2})\times\ldots\times\nonumber\\
&\underbrace{ \sum_{\vec{c_{n-2}}}W(\vec{c_{n-3}};\vec{c_{n-2}})m_{n-1 \rightarrow n-2}(\vec{c_{n-2}})}_{m_{n-2 \rightarrow n-3}(\vec{c_{n-3}})}
\end{align}
\vdots\\
\vdots\\
\begin{equation}
P(\vec{a},\vec{c_1},\vec{b})=\frac{1}{Z}W(\vec{a};\vec{c_1})\underbrace{\sum_{\vec{c_2}}W(\vec{c_1};\vec{c_2})m_{3\rightarrow 2}(\vec{c_2})}_{m_{n-2 \rightarrow n-3}(\vec{c_{n-3}})}
\end{equation}
\end{subequations}
\begin{equation}
P(\vec{a},\vec{b})=\frac{1}{Z}\underbrace{\sum_{\vec{c_1}}W(\vec{a};\vec{c_1})m_{2\rightarrow 1}(\vec{c_1})}_{m_{1 \rightarrow a}(\vec{a})}
\end{equation}
\begin{equation}
\bra{\vec{a}}\rho\ket{\vec{b}}=P(\vec{a},\vec{b})
\end{equation}
In fact the basic idea of this method is that the Hamiltonian  is converted  into the product of smaller statements in such a way that becomes such that these new matrix elements are approximately commutative. When all statements are commutative, the classical BP can be used to solve the problem.\\
\section{Comparison of quantum states:}
Consider two quantum states $\rho$ and $\sigma$. In order to evaluate how close these two states are, a measure for the distance between them is defined \cite{Nielsen21}. This measure not only should has an interesting mathematical properties, but it should has a clear conceptual and practical meaning. The distance which is considered for the two quantum states is like this:\\\\
\textbf{Trace Distance:}\\
\begin{equation}
D(\rho,\sigma)=\frac{1}{2}tr\mid \rho - \sigma \mid
\end{equation}\\
while\\
\begin{equation}
\mid A \mid=\sqrt{A^\dagger A}
\end{equation}\\
it is simply observed that D truly has the properties of a distance. It means that:\\
\begin{equation}
D(\rho,\sigma)=D(\sigma,\rho)
\end{equation}
\begin{equation}
D(\rho,\sigma)=0\leftrightarrow\rho=\sigma
\end{equation}\\
\textbf{Fidelity:}\\
Fidelity between two quantum states gives the proximity degree of the two states. The similarity between the two states of $\rho$ and $\sigma$ is defined by the following relation:\\
\begin{equation}
F(\rho,\sigma):=tr\sqrt{\rho^\frac{1}{2}\sigma\rho^\frac{1}{2}}
\end{equation}
We cannot easily understand by this definition that the fidelity is a symmetric property, ie. $F (\rho, \sigma) = F (\sigma, \rho)$. However, despite its appearance, fidelity is a symmetric property. Some fidelity properties can be soon found by a very little calculation, such that the maximum fidelity is equal to one.\\
\section{Investigation of Heisenberg 3-spin chain:}
ST method and QBP are applied for a Heisenberg 3-spin chain with a Hamiltonian in the form of Eq. 28:\\
\begin{subequations}
\begin{equation}
h_{12}=(S^{x}\otimes S^{x})\otimes I + (S^{y}\otimes S^{y}) \otimes I+(s^{z}\otimes s^{z})\otimes I
\end{equation}
\begin{equation}
h_{23}=I\otimes(S^{x}\otimes S^{x}) + I\otimes(S^{y}\otimes S^{y})+ I\otimes(s^{z}\otimes s^{z})
\end{equation}
\begin{equation}
H\equiv h_{12}+h_{23}
\end{equation}
\end{subequations}
$s^{i}$'s are Pauli matrices.\\
We apply the BP algorithm for the Hamiltonian presented in the above section. The initial messages are selected as follows:
\begin{equation}
   m_{2\rightarrow1}=m_{1\rightarrow2}=m_{2\rightarrow3}=m_{3\rightarrow2}=
  \left[ {\begin{array}{cc}
   1 & 0 \\
   0 & 1 \\
  \end{array} } \right]
\end{equation}
Approximate values of $\rho_{12}$ for Heisenberg Hamiltonian is calculated using ST and QBP approximations for different values of $\beta$. Approximate values are represented by $Q$ in order not to be confused with the accurate values which are assumed to be the result of ST approximation with $n\geq100$. $Q_{12}$ is calculated using ST approach with $n=20$ and QBP. Fidelity and trace distance between approximate($Q's$) and accurate values($\rho's$) are shown in the figures $3$ and $4$, respectively. As it can be seen, approximate values are in very good agreement with the accurate ones. Whatever the $\beta$ is higher, quantum properties in the system becomes bigger. To improve the performance of the ST method, n needs to be increased.\\
Now we want have a debate on complexity, although it is not technical\cite{Cormen22}. What we expect is that with increasing n in the ST approximation, the values obtained are more accurate. What we see in figures $3$ and $4$ for $n = 20$ indicates the same point. The performance of ST method seems to be better than QBP for similar problems.\\
In the example we considered for QBP Algorithm, chain is one-dimensional(fig2). This means that each particle is in a neighborhood of two particles.
\begin{figure}
\includegraphics[width=0.6\textwidth]{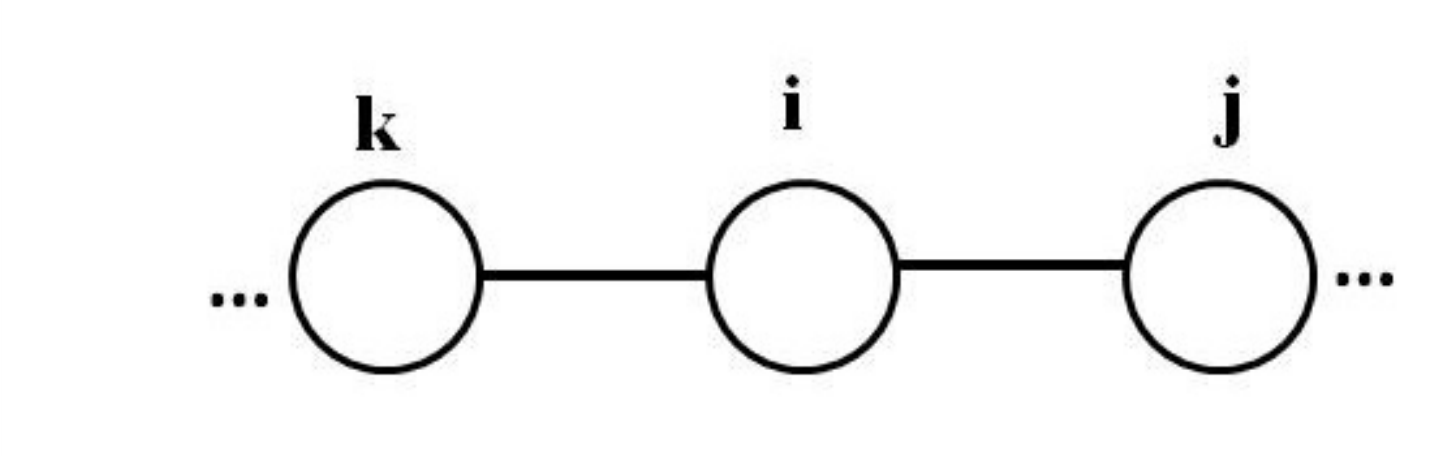}
\caption{one-dimensional spin chain}
\label{fig:2}
\end{figure}
The maximum amount of complexity corresponds to that manner in which all transmitted messages begin from the first particle, after updating rules on them reach to the last particle successively. If $n$ is the number of particles, the number of transmitted messages will be $n-1$, then according to Eq.$(13)$ complexity of this manner can be calculated as follow:\\
\begin{align}
&C(Q_{ij})\propto(n-1)\times2\times2\times2\times2+(n-1)\times2\times2\times2\times2\nonumber\\
&+(n-2)\times4\times4+(n-2)\times4\times4+4\times4=16(4n-5)=2^{6}n-80\nonumber\\
&Time\ Complexity(Q_{ij})\propto O(n)
\end{align}\\
If every message is updated in a loop with order of $m$, then we can write the complexity as follows:\\
\begin{equation}
O(Q_{ij})\propto O(mn)
\end{equation}\\
If we assume that $m$ is approximately equal to $n$ (since the graph is a tree and the order of updating is roughly equal with tree length) the complexity is proportional to $O(n^2)$.\\
For calculating the complexity of ST algorithm, at first we calculate the middle term of this summation as follow:\\
\begin{equation}
\bra{\vec{c_{i-1}}}\exp \lgroup -\frac{\beta}{n}h_{12}\rgroup\exp \lgroup -\frac{\beta}{n}h_{23}\rgroup\ket{\vec{c_{i}}}
\end{equation}\\
\begin{equation}
O(middle\ terms)\propto1\times2^m\times2^m+2^m\times2^m\times1+2^m=2^{m}(2^{m+1}+1)
\end{equation}\\
Then for the first and last terms we have:\\
\begin{align}
&\Braket{\vec{a}|\vec{c_{1}}}\exp \lgroup -\frac{\beta}{n}h_{12}\rgroup\exp \lgroup -\frac{\beta}{n}h_{23}\rgroup\ket{\vec{c_{2}}}\nonumber\\
&\bra{\vec{c_{n-1}}}\exp \lgroup -\frac{\beta}{n}h_{12}\rgroup\exp \lgroup -\frac{\beta}{n}h_{23}\rgroup\ket{\vec{b}}\bra{\vec{b}}
\end{align}\\
\begin{align}
&O(first\ \&\ last\ terms)\propto1\times2^{m}\times2^{m}+1\times2^{m}\times2^{m}+1\times2^{m}\times1+1\times1\times2^{m}\nonumber\\
&=2^{m+1}(2^{m}+1)
\end{align}\\
Finally, the overall complexity can be written in the below form:\\
\begin{align}
&O(Time\ Complexity)\propto2^{m+1}(2^{m}+1)+2^{m+1}(2^{m}+1)+(n-3)2^{m}(2^{m+1}+1)\nonumber\\
&=2^{m+2}(2^{m}+1)+2^{m}(2^{m+1}+1)(n-3)=2^{m}(2^{m+2}+2+(2^{m+1}+1)(n-3))\nonumber\\
&O(Time\ Complexity)\propto O(n2^{m})
\end{align}\\
But what does not appear in computing at first glance is the computational complexity. This complexity is greater in ST approximation, so this is a weak point in comparison with QBP. As it can be seen complexity in ST is exponential, while in QBP is in quadratic form.\\
The complexity of ST approximation enters the problem in a way that to be closer to the accurate values of ST approximation, the value of n needs to be increased. On the other hand, when we look at the form of ST approximation, it seems that we have to calculate the exponential function of an $8\times8$ matrix in an example of a 3-body system. To calculate this exponential function for a 10-body example, we have to calculate the exponential function of a matrix with very large dimensions, which is an extremely challenging issue. A suggestion for diluting this issue is increasing the n. However, as we have described, this increase also creates its own issues. Considering QBP, matrix dimensions are smaller and the number of sums is considerably decreased in compare with ST.QBP is just a basic idea at the beginning and we should be looking for ways to increase its accuracy.
\begin{figure}
\includegraphics[width=0.6\textwidth]{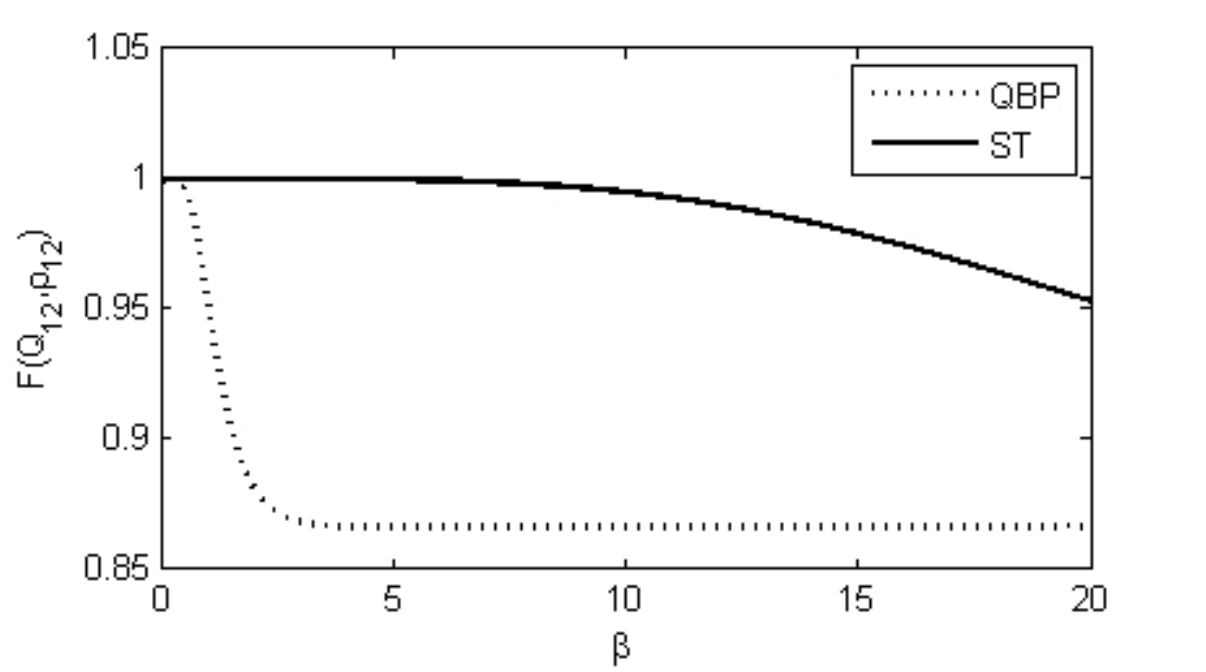}
\caption{Fidelity between approximate values $(Q_{12})$ and accurate ones $(\rho_{12})$; using Suzuki-Trotter approach (solid line) and Quantum Belief Propagation (dashed line)}
\label{fig:3}
\end{figure}
\begin{figure}
\includegraphics[width=0.6\textwidth]{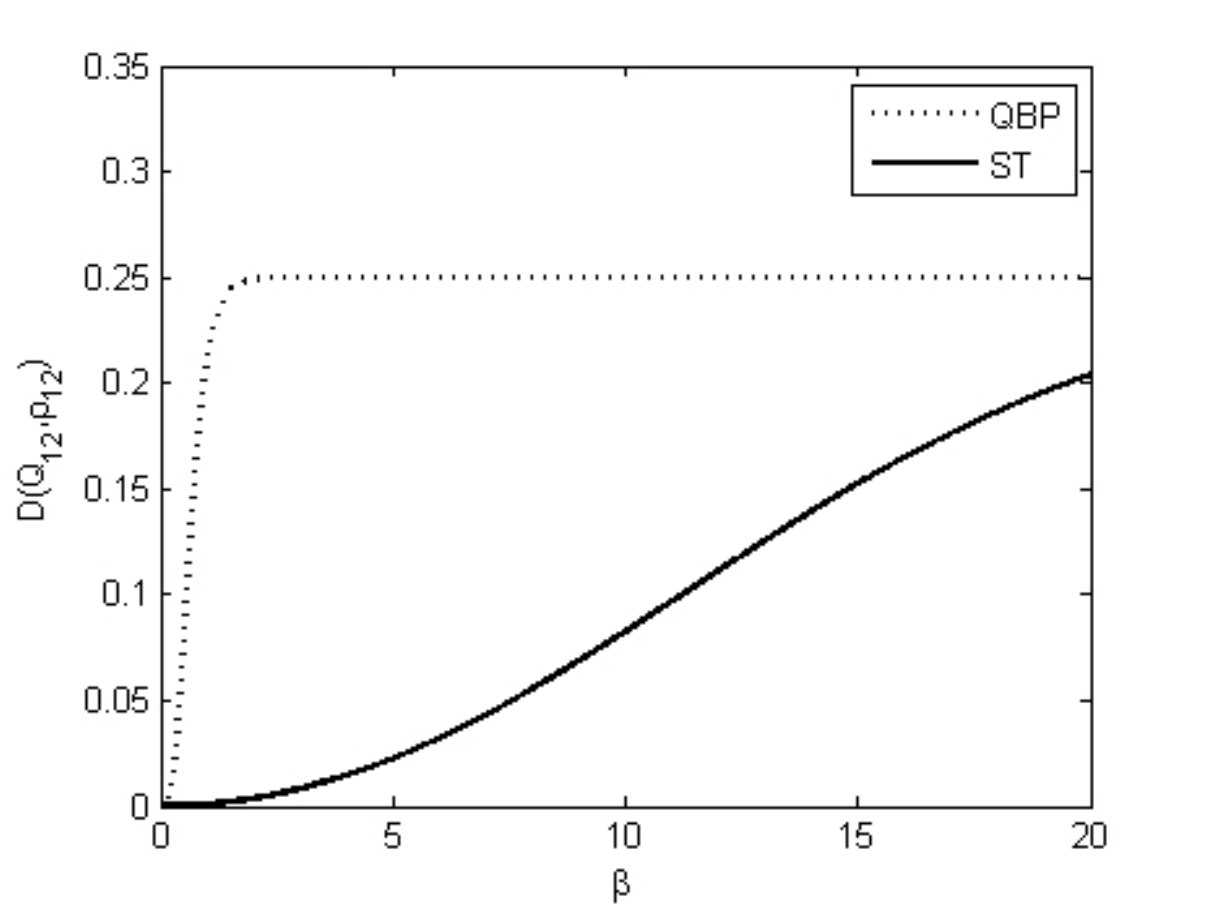}
\caption{Trace distance between approximate values $(Q_{12})$ and accurate ones $(\rho_{12})$; using Suzuki-Trotter approach (solid line) and Quantum Belief Propagation(dashed line)}
\label{fig:4}
\end{figure}\\
\section{Conclusion}
At the end what was done in this article will be described as follows. In multi-body problems, semi-classical Suzuki -Trotter approximation is very accurate in large n. However, there is a high computational complexity in large n and powerful computational tools are needed for calculation. QBP algorithm is less accurate; however, it is far faster than Suzuki -Trotter approximation. Depending on the problem, either of these two approximations can be chosen.

\end{document}